\documentclass[aps,pra,showpacs,superscriptaddress,preprint]{revtex4}

\usepackage{graphicx}
\begin{document}
\title{Effective Mass Schr\"{o}dinger Equation via Point Canonical Transformation}

\author{\small Altuð Arda}
\email[E-mail: ]{arda@hacettepe.edu.tr}\affiliation{Department of
Physics Education, Hacettepe University, Ankara 06800,Turkey}
\author{\small Ramazan Sever}
\email[E-mail: ]{sever@metu.edu.tr}\affiliation{Department of
Physics, Middle East Technical  University, Ankara 06531,Turkey}

\begin{abstract}
Exact solutions of effective radial Schr\"{o}dinger equation are
obtained for some inverse potentials by using the point canonical
transformation. The energy eigenvalues and the corresponding wave
functions are calculated by using a set of mass distributions
\end{abstract}
\pacs{03.65.$-$w 03.65.Ge 12.39.Fd}

\maketitle
In quantum mechanics, exactly solvable problems are
suited in the class having shape invariance potential.$^{[1]}$ The
potential in a given class can be converted to one another by using
the so-called point canonical transformation (PCT).$^{[2]}$ In the
PCT approach, which makes it possible to construct a map between the
wave equations written for two different potentials, it is needed to
know the energy eigenvalues, and the corresponding wave functions of
a given potential (reference potential $U(x)$) to obtain the energy
spectra, and wave functions of the other potential (target potential
$V(x)$). The PCT has been used by many authors $^{[3-6]}$ to find
the energy spectrum, and their wave functions of some potentials in
non-relativistic, and relativistic scheme.

Recently, there are some considerable efforts in the literature
about the developments of the PCT approach to extend to the case
of position-dependent mass (PDM) $^{[2, 7-13]}$ and also the PDM
formalism has found a wide application area in the
literature.$^{[14-18]}$ In this Letter, we choose the potentials
$V(\rho)=-\frac{a}{\rho}+c\rho\,^{p}(p=0, -2)$ and
$V(\rho)=\frac{a}{\rho^2}+c\rho\,^{2}$ as the reference potential
in the case of the radial effective-mass Schr\"{o}dinger equation
(SE), and find the exact energy eigenvalues, and their wave
functions of the target potentials by using the PCT. We follow the
approach studied in Ref.\,[2] where the radial effective mass SE
was solved for isotropic oscillator. We try to show that the above
mentioned formalism could be used for inversely linear and
inversely quadratic potential forms.

The SE is written in the case of PDM ($m_0=\hbar=1$) $^{[2]}$
\begin{eqnarray}
\Big\{\frac{1}{2m(x)}\frac{d^2}{dx^2}-\frac{1}{2m^2(x)}\frac{dm(x)}{dx}\frac{d}{dx}-V(x)\Big\}\phi_{n}(x)
=-E_{n}\phi_{n}(x)\,,
\end{eqnarray}
and in the case of constant mass
\begin{eqnarray}
\Big\{\frac{1}{2}\frac{d^2}{dy^2}-U(y)\Big\}\psi_n
(y)=-\epsilon_{n}\psi_{n}(y),
\end{eqnarray}
where $E_{n}$ and $\phi_{n}(x)$ are the eigenvalues and the
eigenfunctions of the target potential, while $\epsilon_{n}$ and
$\psi_{n}(y)$ denote the eigenvalues, and the eigenfunctions of
the reference potential.

By using the transformations $^{[2]}$
\begin{eqnarray}
y&=&h(x)\,,\\
\psi_{n}(y)&=&\pi(x)\phi_{n}(x).
\end{eqnarray}
from Eq. (2) we obtain
\begin{eqnarray}
\Big\{\frac{d^2}{dx^2}&+&f_{1}(\pi(x);h(x))\frac{d}{dx}+f_{2}(\pi(x);h(x))\nonumber\\&-&
2[h'(x)]^2\Big[U(h(x))-\epsilon_{n}\Big]\Big\}\psi_{n}(x)=0\,,
\end{eqnarray}
where
\begin{eqnarray}
f_{1}(\pi(x);h(x))&=&2\pi'(x)\Bigg[\,\frac{1}{\pi(x)}-\frac{h''(x)}{\pi'(x)h'(x)}\,\Bigg]\,,\\
f_{2}(\pi(x);h(x))&=&\pi''(x)\Bigg[\,\frac{1}{\pi(x)}-\frac{h''(x)\pi'(x)}{\pi(x)\pi''(x)h'(x)}\,\Bigg]\,,
\end{eqnarray}
and prime denotes the derivative to the spatial coordinate. If the
following equalities are satisfied,
\begin{eqnarray}
h'(x)&=&m(x)\pi^2(x)\,,\\
V(x)-E_{n}&=&\frac{[h'(x)]^2}{m(x)}\Big[U(h(x))-\epsilon_{n}\Big]+f(m(x);h(x))\,,
\end{eqnarray}
then the transformation given in Eqs. (3) and (4) is a point
canonical transformation.$^{[2]}$ The function $f(m(x);h(x))$ in
the above equation is given
\begin{eqnarray}
f(m(x);h(x))=\frac{1}{4m(x)}\Bigg[\frac{m''(x)}{m(x)}-\frac{3}{2}\Bigg(\frac{m'(x)}{m(x)}\Bigg)^2-
\frac{h'''(x)}{h'(x)}+\frac{3}{2}\Bigg(\frac{h''(x)}{h'(x)}\Bigg)^2\Bigg].
\end{eqnarray}
Taking $h'(x)=\sqrt{m(x)}$ and defining the parameter $\sigma(x)$
as $\sigma(x)=\frac{1}{\beta}\int h'(x)dx$, we obtain the energy
spectrum, the corresponding wave functions, and the target
potential as
\begin{eqnarray}
E_{n}&=&\epsilon_{n}\,,\\
\phi_{n}(x)&=&\sqrt{\frac{h'(x)}{m(x)}}\psi_{n}(\beta\sigma(x)),\\
V(x)&=&U(\beta\sigma(x))+\frac{1}{4m(x)}\Bigg[\frac{1}{2}\frac{m''(x)}{m(x)}-\frac{7}{8}\Bigg(\frac{m'(x)}{m(x)}
\Bigg)^2\Bigg].
\end{eqnarray}
where the parameter $\beta$ is the scale parameter.$^{[2]}$

Giving a PDM function and taking an exactly solvable reference
potential $U(y)$, we construct a PCT defined by Eqs. (3) and (4)
and obtain a target potential, which can be used to solve the
effective mass SE exactly, eigenvalues, and wave functions of the
potential.

Now, let us look the case of the radial effective-mass SE, which
can be written as $^{[2]}$
\begin{eqnarray}
\Big\{\frac{1}{2m(r)}\frac{d^2}{dr^2}-\frac{\ell(\ell+1)}{2m(r)r^2}+\frac{m'(r)}{2m^2(r)}\Bigg(\frac{1}{r}-
\frac{d}{dr}\Bigg)-V(r)\Big\}\phi_{n\ell}(r)=-E_{n\ell}\phi_{n\ell}(r),
\end{eqnarray}
where $\ell$ is the angular-momentum quantum number and the radial
wave function is taken as $\Psi_{n\ell}(r)=\phi_{n\ell}(r)/r$.

The radial SE for the case of constant mass reads
\begin{eqnarray}
\Big\{\frac{1}{2}\frac{d^2}{d\rho^2}-\frac{\ell'(\ell'+1)}{\rho^2}-U(\rho)\Big\}\psi_{n\ell'}(r)=-\epsilon_{n\ell'}
\psi_{n\ell'}(r).
\end{eqnarray}
where $\ell'$ is the angular-momentum. By using the
transformations
\begin{eqnarray}
\rho&=&h(r),\\
\psi_{n\ell'}(r)&=&\pi(r)\phi_{n\ell}(r).
\end{eqnarray}
from Eq.\,(15) we obtain
\begin{eqnarray}
\Big\{\frac{d^2}{dr^2}&+&f_{1}(\pi(r);h(r))\frac{d}{dr}+f_{2}(\pi(r);h(r))
-\ell'(\ell'+1)\Bigg(\frac{h'(r)}{h(r)}\Bigg)^2\nonumber\\&-&
2[h'(r)]^2\Big[U(h(r))-\epsilon_{n\ell'}\Big]\Big\}\psi_{n\ell'}(r)=0,
\end{eqnarray}
Comparing with Eq. (14) gives $\pi(r)=\sqrt{h'(r)/m(r)}$ and the
following
\begin{eqnarray}
V(r)-E_{n\ell}+\frac{\ell(\ell+1)}{2m(r)r^2}&=&\frac{[h'(r)]^2}{m(r)}\Big[U(h(r))-
\epsilon_{n\,\ell'}\Big]+\frac{\ell'(\ell'+1)}{2m(r)}\Bigg(\frac{h'(r)}{h(r)}\Bigg)^2\nonumber\\&+&
\frac{m'(r)}{2m^2(r)r}+f(m(r);h(r))\,.
\end{eqnarray}
which gives the target potential $V(r)$ and its energy eigenvalues
$E_{n\ell}$ for a given radial effective mass function
$m(r)$.$^{[2]}$

\textit{A) Kratzer potential.}\\
We consider the following form of the one-dimensional Kratzer
potential$^{[19]}$
\begin{eqnarray}
U(y)=\frac{A}{y}+\frac{B}{y^2}\,,
\end{eqnarray}
which has been extensively used to explain the molecular structure
and interactions in quantum and molecular chemistry.$^{[20]}$ The
energy eigenvalues and the corresponding wave functions of the
reference potential $U(y)$ are$^{[21]}$
\begin{eqnarray}
\epsilon_{n\ell'}&=&-\frac{2A^2}{[2n+1+\sqrt{1-16B}\,]^2},\\
\psi_{n\ell'}(r)&=&a_{n}\,y^{\frac{1}{2}\,+\sqrt{-A}}e^{-\xi y}
\,_1F_1(-n;1+2\sqrt{-A};2\xi y);\xi=\sqrt{-2\epsilon_{n\ell'}},
\end{eqnarray}
where $\,_1F_1(-n;\varrho;x)$ are the hypergeometric type
function. Now, we give the explicit form of the target potentials
and its wave functions for the following three different mass
distributions.

\texttt{(1) Mass function $m(x)=(\delta+x^2)^2/(1+x^2)^2$} We
obtain $\sigma(x)$ as $\sigma(x)=(1/\tau)[x+(\delta-1)/tan\,x]$
and the target potential from Eq.\,(13)
\begin{eqnarray}
V(x)=\frac{\theta\beta}{\Big[x+\frac{\delta-1}{tan\,x}\Big]}\Bigg[1+\frac{\theta\beta}{x+\frac{\delta-1}{tan\,x}}
\Bigg]+\frac{\delta-1}{2}\frac{3x^4+
2(2-\delta)x^2-\delta}{(\delta+x^2)^4}\,,
\end{eqnarray}
where $\theta=A/\beta$ and we have set the potential parameter $B$
as $B=A^2$. The energy spectrum and the corresponding wave
functions are written from Eqs.\,(11) and (12) as
\begin{eqnarray}
E_{n}&=&-\frac{\theta^2\beta^2}{[2n+1+\sqrt{1-16\theta^2\beta^2}]^2},\\
\phi_{n}(x)&=&a_{n}\sqrt{\frac{\delta+x^2}{1+x^2}}[\beta\sigma(x)]^{\frac{1}{2}\,+\sqrt{-A}}\,
exp\,[-\xi\beta\sigma(x)]
_1F_1(-n;1+2\sqrt{-A};2\xi\beta\sigma(x))\,.
\end{eqnarray}
\texttt{(2) Mass function $m(x)=a/(\delta'+x^2)$} The distribution
corresponds to an asymptotically vanishing mass function. It gives
$\sigma(x)=(\sqrt{a}/\beta)ln\,[x+\sqrt{1+x^2}]$. Therefore, we
obtain the target potential
\begin{eqnarray}
V(x)=\frac{\theta\beta}{\sqrt{a\,}\,ln\,[x+\sqrt{1+x^2}]}\Bigg[1+\frac{\theta\beta}
{\sqrt{a}\,ln\,[x+\sqrt{1+x^2}\,]}\Bigg]
-\frac{x^2+2\delta'}{8a(\delta'+x^2)},
\end{eqnarray}
and the energy spectra and their wave functions are
\begin{eqnarray}
E_{n}&=&-\frac{\theta^2\beta^2}{[2n+1+\sqrt{1-16\theta^2\beta^2}\,]^2}\,,\\
\phi_{n}(x)&=&\frac{A_{n}}{(\delta'+x^2)^{1/4}}\,[\beta\sigma(x)]^{\frac{1}{2}\,+\sqrt{-A}}\,
exp\,[-\xi\beta\sigma(x)]_1F_1(-n;1+2\sqrt{-A};2\xi\beta\sigma(x))\,.
\end{eqnarray}
where the normalization constant is $A_n=a^{1/4}\,a_n$.

\texttt{(3) Mass function $m(x)=1+tanh(\delta'' x)$} We obtain the
following target potential, energy eigenvalues and the
corresponding eigenfunctions for this case,
\begin{eqnarray}
V(x)&=&\theta[\sigma(x)]^{-1}\Big[1+\theta[\sigma(x)]^{-1}\Big]-\frac{(\delta'')^2}{32}
\frac{7+tanh(\delta''x)}{[sinh(\delta''x)+cosh(\delta''x)]^2}\,,\\
E_{n}&=&-\frac{\theta^2\beta^2}{[2n+1+\sqrt{1-16\theta^2\beta^2}]^2}\,,\\
\phi_{n}(x)&=&a_{n}\Big[1+tanh(\delta''x)\Big]^{1/4}[\beta\sigma(x)]^{\frac{1}{2}\,+\sqrt{-A}}\nonumber\\
&\times&exp[-\xi\beta\sigma(x)]
_1F_{1}(-n;1+2\sqrt{-A};2\xi\beta\sigma(x)),
\end{eqnarray}
where
$\sigma(x)=[\sqrt{2}/(\beta\delta'')]tanh^{-1}[\sqrt{1+tanh(\delta''x)}/\sqrt{2}\,]$\,.
It is seen that the Schrödinger equation can be exactly solved by
using point canonical transformation in one-dimension.

\textit{(B) Family of power law plus inverse power potentials.} We
study the target potential and its energy eigenvalues, and wave
functions for the following reference potentials in the form
$U(\rho)=-\frac{a}{\rho}+c\rho^{p} \,(p=0, -2)$ and
$U(\rho)=\frac{a}{\rho^2}+c\rho\,^{2}$ for the radial SE for the
case of position dependent mass.

\textsf{(1) $U(\rho)=-\frac{a}{\rho}+c$.} The reference potential
has the form
\begin{eqnarray}
U(\rho)=-\frac{a}{\rho}+c\,,
\end{eqnarray}
which describes a hydrogen atom with the energy eigenvalue $E-c$
having the energy eigenvalues and corresponding wave functions,
\begin{eqnarray}
\epsilon_{n\ell'}&=&c-\frac{a^{2}}{2}(n+\ell'+1)^{-2}\,,\\
\psi_{n\ell'}(\rho)&=&a_{n}\rho^{\ell'+1}exp\Big[-\frac{a}{n+\ell'+1}\rho\Big]
\,_{1}F_{1}(-n;2\ell'+2;\frac{2a}{n+\ell'+1}\,\rho)\,.
\end{eqnarray}
Taking the radial mass function $m(r)=\mu r^{\kappa}$ and the PCT
function $h(r)=r^{\nu}$ and inserting into Eq.\,(19), we obtain
\begin{eqnarray}
V(r)-E_{n\ell}+\frac{\ell(\ell+1)}{2\mu}r^{-\kappa-2}&=&-\frac{a\nu^2}{\mu}
r^{\nu-\kappa-2}-\frac{\nu^2\epsilon_{n\ell'}}{\mu}r^{2\nu-\kappa-2}
\nonumber\\&+&\frac{r^{-\kappa-2}}{2\mu}\Big[\ell'(\ell'+1)\nu^2+\kappa+\frac{1}{2}\kappa(\kappa-1)\nonumber\\
&-&\frac{3}{4}\kappa^2-\frac{1}{2}(\nu-1)(\nu-2)+\frac{3}{4}(\nu-1)^2\Big]\,,
\end{eqnarray}
which gives two possible solutions

(a) For the case of $\nu=1+\frac{\kappa}{2};\kappa\neq-2$, from
Eq.\,(35) we obtain
\begin{eqnarray}
V(r)&=&-\mu C^2r^{-1-\frac{\kappa}{2}}\,,\\
E_{n\ell}&=&-\frac{1}{2}\,a\mu
C^2\Big(n+|\mathcal{L}(\ell)|+\frac{1}{2}\Big)^{-2}\,,
\end{eqnarray}
where
$|\mathcal{L}(\ell)|=\mp\frac{1}{\kappa+2}\,\sqrt{4\ell(\ell+1)+(\kappa-1)^2\,}$
and $C$ is a real coupling parameter. From Eq.\,(17) we obtain
corresponding wave functions
\begin{eqnarray}
\phi_{n\ell}(r)&=&b_{n}\sqrt{\frac{2\mu}{\kappa+2}}exp\Big[-\zeta
r^{1+\frac{\kappa}{2}}\Big]r^{(1+\frac{\kappa}{2})
(|\mathcal{L}(\ell)|+\frac{1}{2})+\frac{\kappa}{4}}\nonumber\\
&\times& _1F_{1}(-n;2|\mathcal{L}(\ell)|+1;2\zeta
r^{1+\frac{\kappa}{2}}),
\end{eqnarray}
where $\zeta=a'/(n+|\mathcal{L}(\ell)|+\frac{1}{2})$, and $b_{n}$
is a normalization constant.

(b) For the case of $\nu=\kappa+2;\kappa\neq-2$, following the
above procedure, we obtain the target potential, the eigenvalues
and the eigenfunctions by taking into account the independence of
a potential from any quantum number,
\begin{eqnarray}
V(r)&=&-\frac{1}{2}\mu Cr^{\kappa+2}\,,\\
E_{n\ell}&=&\frac{\mu C}{2a}\Big(2n+2\mathcal{L}(\ell)+1\Big)^{2},
\end{eqnarray}
\begin{eqnarray}
\phi_{n\ell}(r)&=&b_{n}\sqrt{\frac{\mu}{\kappa+2}}exp\Big[-\zeta
r^{\kappa+2}\Big]r^{(\kappa+2)
(|\mathcal{L}(\ell)|+\frac{1}{2})-\frac{1}{2}}\nonumber\\
&\times& _1F_{1}(-n;2|\mathcal{L}(\ell)|+1;2\zeta
r^{1+\frac{\kappa}{2}}),
\end{eqnarray}
where $b_{n}$ is a normalization constant and $C$ is the coupling
parameter .$^{[2]}$

The particular case $\kappa=-2$ is solved by taking
$h(r)=\frac{1}{a}\,ln(r)$$^{[2]}$ to obtain the target potential.
In that case the energy eigenvalues, and the corresponding wave
functions only exist for the ground state.$^{[2]}$ We summarize
the potential and the energy eigenvalues by using Eq.\,(19)
\begin{eqnarray}
V(r)&=&-\frac{1}{\mu ln(r)}+\frac{1}{2}\,Cln^{-2}(r)\,,\\
E_{n\ell}&=&\frac{1}{\mu}\Big\{-\frac{1}{2}\Big[n+\frac{1}{2}+|\Lambda(\ell)|\Big]^{-2}-\frac{9}{8}\Big\},
\end{eqnarray}
where $|\Lambda(\ell)|=\pm\frac{1}{2}\sqrt{1+4\mu C}$\,. The
corresponding wave functions are
\begin{eqnarray}
\phi_{n\ell}(r)=b_{n}r^{-1/2}[ln(r)]^{\frac{1}{2}+|\Lambda(\ell)|}\,exp\,[-\eta
ln(r)] _1F_{1}(-n;1+2|\Lambda(\ell)|;2\eta ln(r)),
\end{eqnarray}
where $\eta=(n+\frac{1}{2}+|\Lambda(\ell)|)^{-1}$ and
$b_{n}=\sqrt{\mu}(a')^{-|\Lambda(\ell)|}$\,.

\textsf{(2)\,$U(\rho)=-\frac{a}{\rho}+\frac{c}{\rho^{2}}$.} The
reference potential has the form
\begin{eqnarray}
U(\rho)=-\frac{1}{\rho}(a+\frac{c}{\rho})\,,
\end{eqnarray}
and the energy eigenvalues and the corresponding wave functions
are listed as $^{[22]}$
\begin{eqnarray}
\epsilon_{n\ell'}&=&-\frac{a^2}{2}\bigg(n+\frac{1}{2}+\sqrt{(\ell'+\frac{1}{2})^2+2c}\bigg)^{-2}\,,\\
\psi_{n\ell'}(\rho)&=&a_{n}exp\,[-\frac{a}{n+\Lambda+1}\rho]
_1F_{1}(-n;2\Lambda+2;\frac{2a}{n+\Lambda+1}\rho),
\end{eqnarray}
where $\Lambda=-\frac{1}{2}+\sqrt{(\ell'+\frac{1}{2})^2+2c}$.
Following the same procedure, we can write the results of two
possible solutions

(a) For the case of $\nu=1+\frac{\kappa}{2};\kappa\neq-2$, the
target potential and its energy spectrum are written as
\begin{eqnarray}
V(r)&=&-\frac{a\gamma^2}{\mu}r^{-1-\frac{\kappa}{2}}\,,\\
E_{n\ell}&=&-\frac{a^2\gamma^2}{2\mu}\bigg(n+\frac{1}{2}+\sqrt{|\mathcal{L}(\ell)|^2+2c\,}\bigg)^{-2},
\end{eqnarray}
where
$\mathcal{L}(\ell)=\mp\frac{1}{\kappa+2}\sqrt{4\ell(\ell+1)+(\kappa-1)^2-2c(\kappa+2)^2}$
and the corresponding wave functions read
\begin{eqnarray}
\phi_{n\ell}(r)&=&b_{n}\sqrt{\frac{2\mu}{\kappa+2}}r^{\kappa/4}
exp\,[-\frac{a}{n+\frac{1}{2}+\sqrt{|\mathcal{L}(\ell)|^2+2c}}r^{1+\frac{\kappa}{2}}]\nonumber\\
&\times&_1F_{1}
(-n;1+2\sqrt{|\mathcal{L}(\ell)|^2+2c};\frac{2a}{n+\frac{1}{2}+\sqrt{|\mathcal{L}(\ell)|^2+2c}}
r^{1+\frac{\kappa}{2}}).
\end{eqnarray}
where $b_{n}$ is a normalization constant.

(b)For the case of $\nu=\kappa+2;\kappa\neq-2$, we list the target
potential and the energy spectrum keeping in mind that any
potential function must be independent from any quantum number
\begin{eqnarray}
V(r)&=&-2\mu Cr^{\kappa+2}\,,\\
E_{n\ell}&=&-\frac{4\mu
C}{a}\frac{1}{\bigg(n+\frac{1}{2}+\sqrt{|\mathcal{L}(\ell)|^2+2c}\bigg)^{-2}}\,,
\end{eqnarray}
where $\mathcal{L}(\ell)$ is given in $(a)$. The corresponding
wave functions are written as
\begin{eqnarray}
\phi_{n\ell}(r)&=&b_{n}\sqrt{\frac{\mu}{\kappa+2}}r^{-1/2}
exp\,[-\frac{a}{n+\frac{1}{2}+\sqrt{|\mathcal{L}(\ell)|^2+2c}}]\nonumber\\
&\times&_1F_{1}
(-n;1+2\sqrt{|\mathcal{L}(\ell)|^2+2c\,};\frac{2a}{n+\frac{1}{2}+\sqrt{|\mathcal{L}(\ell)|^2+2c}}
r^{\kappa+2})\,.
\end{eqnarray}
Finally, we study the particular case where $\kappa=-2$ by taking
$h(r)=\frac{1}{a}\,ln(r)$\,. By using Eq.\,(19), we summarize the
target potential and its energy eigenvalues as
\begin{eqnarray}
V(r)&=&-\frac{1}{\mu ln(r)}+\frac{1}{ln^{2}(r)}\big[\frac{c}{\mu}+\frac{C}{2}\big]\,,\\
E_{n\ell}&=&\frac{1}{\mu}\Big\{-\frac{1}{2}\Big[n+\frac{1}{2}+\sqrt{|\Lambda'(\ell)|+2c}\Big]^{2}-\frac{9}{8}\Big\},
\end{eqnarray}
where $|\Lambda'(\ell)|=\pm\frac{1}{2}\sqrt{1+4\mu C}$. The
corresponding wave functions are
\begin{eqnarray}
\phi_{n\ell}(r)=b_{n}r^{-1/2}exp\,[-\eta' ln(r)]
_1F_{1}(-n;1+2\sqrt{|\Lambda'(\ell)|^2+2c};2\eta' ln(r)),
\end{eqnarray}
where
$\eta'=\bigg(n+\frac{1}{2}+\sqrt{|\Lambda'(\ell)|^2+2c}\bigg)^{-1}$
and $b_{n}=\sqrt{a\mu}$.

\textsf{(3) $U(\rho)=\frac{a}{\rho^2}+c\rho\,^{2}$.} The energy
eigenvalues and corresponding wave functions of this potential can
be written as$^{[23]}$
\begin{eqnarray}
\epsilon_{n\ell'}&=&\sqrt{2c}\bigg[2n+1+\frac{1}{2}\sqrt{(2\ell'+1)^2+8a}\bigg],\\
\psi_{n\ell'}(\rho)&=&(2c)^{\delta/2}\rho^{2\delta}exp\bigg[-\frac{c}{2}\rho^2\bigg]\
_1F_1(-n,2\delta+\frac{1}{2};\sqrt{2c\,}\rho^2)\,.
\end{eqnarray}
where $\delta=\frac{1}{4}[1+\sqrt{(2\ell'+1)^2+8a}]$. In this
case, we summarize the two possible solutions as

(a) For the case of$\nu=1+\frac{\kappa}{2};\kappa\neq-2$,
\begin{eqnarray}
V(r)&=&\frac{c\nu^2}{\mu}\,r^{\kappa+2},\\
E_{n\ell}&=&\sqrt{\frac{2c\nu^4}{\mu^2}}\bigg(2n+1+\sqrt{2|\mathcal{L}(\ell)|^2+2a}\bigg),
\end{eqnarray}
where
$\mathcal{L}(\ell)=\mp\frac{1}{\kappa+2}\sqrt{4\ell(\ell+1)+(\kappa-1)^2-2a(\kappa+2)^2}$,
and the corresponding wave functions
\begin{eqnarray}
\phi_{n\ell}(r)=b'_{n}\sqrt{\frac{2\mu}{\kappa+2}}r^{\delta(\kappa+2)+\frac{\kappa}{4}}
exp\bigg[-\sqrt{\frac{c}{2}}r^{\kappa+2}\bigg]\,
_1F_{1}\bigg(-n,2\delta+\frac{1}{2};\sqrt{2c}r^{\kappa+2}\bigg),
\end{eqnarray}
where $b'_{n}=b_{n}(2c)^{\delta/2}$ and $b_{n}$ is a normalization
constant.

(b) For the case
of$\nu=\frac{1}{2}+\frac{\kappa}{4};\kappa\neq-2$, we give the
target potential and its energy eigenvalues as
\begin{eqnarray}
V(r)&=&-2\mu Cr^{-\frac{\kappa}{2}-1},\\
E_{n\ell}&=&-\sqrt{2C^2\mu^2c}\frac{1}{2n+1+\sqrt{|\mathcal{L'}(\ell)|^2+2a}},
\end{eqnarray}
where
$\mathcal{L'}(\ell)=\mp\frac{1}{\kappa+2}\sqrt{16\ell(\ell+1)+4(\kappa-1)^2-2a(\kappa+2)^2\,}$.
The corresponding wave functions are written as
\begin{eqnarray}
\phi_{n\ell}(r)=b'_{n}\sqrt{\frac{2\mu}{\kappa+2}}r^{\delta(1+\frac{\kappa}{2})+\frac{\kappa}{4}}
exp\bigg[-\sqrt{\frac{c}{2}}r^{1+\frac{\kappa}{2}}\bigg]
\,_1F_{1}\bigg(-n,2\delta+\frac{1}{2};\sqrt{2c}r^{\frac{\kappa}{2}}\bigg).
\end{eqnarray}
Now, we proceed the particular case where $\kappa=-2$ as a last
situation by taking $h(r)=\frac{1}{a}\,ln(r)$\,. Following the
same procedure gives
\begin{eqnarray}
V(r)&=&\frac{c}{\mu
a^4}\,ln^{2}(r)+\frac{1}{ln^{2}(r)}\big(\frac{a}{\mu}+\Theta\big)\,,\\
E_{n\ell}&=&\sqrt{\frac{2c}{\mu^2a^4}}\bigg(2n+1+\sqrt{\left|\Gamma(\ell)\right|^2+2a}\bigg)-\frac{9}{8\mu}.
\end{eqnarray}
where $\Gamma(\ell)=\pm\frac{1}{2}\sqrt{1+8\mu\Theta}$ and
$\Theta$ is a real coupling parameter.$^{[2]}$ The wave functions
can be given as
\begin{eqnarray}
\phi_{n\ell}(r)=b_{n}\sqrt{\mu a
}\bigg(\frac{\sqrt{2c}}{a^2}\bigg)^{\delta}r^{-(1/2)}exp\bigg[-\sqrt{\frac{c}{2a^4}}ln^{2}(r)\bigg]
\,_1F_{1}\bigg(-n,2\delta+\frac{1}{2};\sqrt{\frac{2c}{a^4}}ln^{2}(r)\bigg),
\end{eqnarray}
where
$\delta=\frac{1}{4}[1+\sqrt{4\left|\Gamma(\ell)\right|^2+8a}]$.

In summary, we have used the PCT in the position-dependent mass
formalism and choose the Kratzer-like potential in one-dimension
as the reference potential to construct the exact solvable target
potential and write the energy eigenvalues, and the corresponding
wave functions in closed forms. We apply the extended formalism of
the PCT to the three-dimensional problems and use the potentials
$V(\rho)=-\frac{a}{\rho}+c\rho\,^{p}(p=0, -2)$ and
$V(\rho)=\frac{a}{\rho^2}+c\rho\,^{2}$ as the reference
potentials. The results show that a suitable transformation can be
used to convert the wave equation into the exact solvable SE and
therefore the required energy eigenvalues and wave functions are
obtained exactly.


\begin{thebibliography}{99}



\bibitem{1} Gendenstein L 1983 JETP Lett. {\bf 38} 356



\bibitem{2} Alhaidari A D 2002 Phys. Rev. A {\bf 66} 042116 



\bibitem{3} De R, Dutt R and Sukhatme U 1992 J. Phys. A {\bf 25} L843



\bibitem{4} Dutt R, Khare A and Varshni Y P 1995 J. Phys. A {\bf 28} L107 



\bibitem{5} Jia C S, Diao Y F, Li M, Yang Q B, Sun L T and Huang R Y 2004
J. Phys. A {\bf 37} 11275 


\bibitem{6} Alhaidari A D 2001 Phys. Rev. Lett. {\bf 87} 210405 



\bibitem{7} Roy B and Roy P 2002 J. Phys. A {\bf 35} 3961 


\bibitem{8} Dutra A S and Almeida C A S 2000 Phys. Lett. A {\bf 275} 25 



\bibitem{9} Plastino A R, Rigo A, Casas M, Gracias F and Plastino
A 1999 Phys. Rev. A {\bf 60} 4318 


\bibitem{10} Alhaidari A D 2003 Int. J. Theor. Phys. {\bf 42} 2999 


\bibitem{11} Yu J and Dong S H 2004 Phys. Lett. A {\bf 325} 194 


\bibitem{12} Jiang L, Yi L Z and Jia C S 2005 Phys. Lett. A {\bf 345} 279 


\bibitem{13} Tezcan C and Sever R 2008 Int. J. Theor. Phys. {\bf 47} 1471 




\bibitem{14} Serra L and Lipparini E 1997 Europhys. Lett. {\bf 40} 667 



\bibitem{15} von Roos O 1983 Phys. Rev. B {\bf 27} 7547 



\bibitem{16} von Roos O and Mavromatis H 1985 Phys. Rev. B {\bf 31} 2254 



\bibitem{17} de Saavedra F A, Boronat J, Polls A and Fabrocini A
1994 Phys. Rev. B {\bf 50} 4248 



\bibitem{18} Barranco M, Pi M, Gatica S M, Hernandez E S and Navarro J
1997 Phys. Rev. B {\bf 56} 8997 



\bibitem{19} Kratzer A 1920 Z. Phys. {\bf 3} 289 



\bibitem{20} Roy R J L and Bernstein R B 1970 J. Chem. Phys. {\bf 52} 3869 



\bibitem{21} Maharana K, [arXiv: math-ph/0401026]



\bibitem{22} Aygun M, Bayrak O and Boztosun I 2007 J. Phys. B: At. Mol. Opt. Phys.
{\bf 40} 5372 



\bibitem{23} Constantinescu F and Magyari E 1971 Problems in Quantum Mechanics (Oxford:
Pergamon.)
\end{thebibliography}
\end{document}